\documentclass[preprint,12pt,authoryear]{elsarticle}

\usepackage{amssymb}
\usepackage{tikz-cd}
\usepackage{amsthm, amsmath}

\newtheorem{theorem}{Theorem}
\newtheorem{definition}{Definition}
\newtheorem{lemma}{Lemma}

\journal{Information Processing Letters}

\begin{document}

\begin{frontmatter}



\title{A simple information theoretical proof of the Fueter-P\'{o}lya Conjecture}


\author{Pieter W. Adriaans}

\address{ILLC, FNWI-IVI, SNE\\
University of Amsterdam,\\Science Park 107\\ 1098 XG Amsterdam, \\The
Netherlands.}

\ead{P.W.Adriaans@uva.nl}

\begin{abstract}
We present a simple information theoretical proof of the Fueter-P\'{o}lya Conjecture: there is no polynomial pairing function that defines a bijection between the set of natural numbers $\mathbb{N}$ and its product set $\mathbb{N}^2$ of degree higher than $2$. We introduce the concept of information efficiency of a function as the balance between the information in the input and the output. We show that 1) Any function defining a computable  bijection between an infinite set and the set of natural numbers is information efficient, 2) the Cantor functions satisfy this condition, 3) any hypothetical higher order function defining such a bijection also will be information efficient, i.e. it stays asymtotically close to the Cantor functions and thus cannot be a higher order function.  
\end{abstract}


\end{frontmatter}



\section{Introduction}
\label{}

The set of natural numbers $\mathbb{N}$ can be mapped to its product set by the two so-called Cantor pairing functions $\pi^2: \mathbb{N}^2 \rightarrow \mathbb{N}$ that define a two-way polynomial time computable bijection:

\begin{equation}\label{CANTORPAIRING}
\pi^2(x,y) := 1/2(x + y)(x + y + 1)+y
\end{equation}
 
The Fueter - P\'{o}lya theorem (\cite{FP23}) states that the Cantor pairing function and its symmetric counterpart $\pi'^2(x,y)=\pi^2(y,x)$ are the only possible quadratic pairing functions. The original proof by Fueter and P\'{o}lya is complex, but a simpler version was published in \cite{vsemirnov2002two} (cf. \cite{nathanson2016cantor}). The  Fueter - P\'{o}lya conjecture states that there are no other polynomial functions that define such a bijection. In this paper we present a proof of this conjecture based on the information efficiency of bijections. We introduce the concept of information efficiency of a function as the balance between the information in the input and the information in the output. We show that every computable bijection from a set to the set of natural numbers is information efficient and that the Cantor functions satisfy this constraint. Any other function satisfying this constraint has to stay asymptotically close to the Cantor functions and thus cannot have a higher order. 

\section{Proof}
We use $K(x)$ as the prefix-free Kolmogorov complexity of $x$ (\cite{LiVi08}). The \emph{Information Efficiency} of a function is the difference between the amount of information in the input of a function and the amount of information in the output. We use the shorthand $f(\overline{x})$ for  $f(x_1,x_2,\dots,x_k)$: 

\begin{definition}[Information Efficiency of a Function]\label{EFFFUNCTION}
Let $f: \mathbb{N}^k \rightarrow \mathbb{N}$  be a function of $k$ variables.  We have:
\begin{itemize} 
\item the \emph{input information} $I(\overline{x})$ and 
\item the \emph{output information}   $I(f(\overline{x}))$. 
\item The information efficiency of the expression $ f(\overline{x})$ is  
\[\delta(f(\overline{x}))= I(f(\overline{x})) - I(\overline{x})\]
\item A  function $f$ is \emph{information conserving} if $\delta(f(\overline{x}))=0$ i.e. it contains exactly the amount of information in its input parameters, 
\item it is \emph{information discarding} if  $\delta(f(\overline{x}))<0$ and 
\item it has \emph{constant information } if  $\delta(f(\overline{x})) = c$. 
\item it is \emph{information expanding} if  $\delta(f(\overline{x}))>0$. 
\end{itemize}
\end{definition}

The following theorem describes a fundamental quality of deterministic information processing: 

\begin{theorem}\label{FUND}
No finite deterministic program expands information. 
\end{theorem}
Proof: This result holds for a variety of information measures. We prove the case for Shannon information and Kolmogorov complexity. Suppose $p(i)=o$, where $p$ is a finite deterministic program and $i$ and $o$ are input and output. Applying Shannon's theory  we have $P(o|p,i)=1$, i.e. the occurrence message $o$ given $p$ and $i$ is certain. Consequently $- \log P(o|p,i)=0$, so   $o$ contains $0$ bits of new information. According to Kolmogorov complexity, since the program $p$ has finite length,  $K(o) \leq K(i) + O(1)$. $\Box$ 

\begin{lemma}\label{EFF}
If $f: \mathbb{N}^k \rightarrow \mathbb{N}$ and the corresponding function $f^{-1}$ define a deterministically computable  bijection then $f$ is information efficient on all elements of $\mathbb{N}^k$. 
\end{lemma}
Proof: The functions $f$ and $f^{-1}$ are deterministic programs, so by theorem \ref{FUND} they cannot expand information. But $f$ cannot discard information either. Since $f$ is bijection we have $f^{-1}f(\overline{x}) =  \overline{x}$ and if $f$ is information discarding $f^{-1}$ would be information expanding. $\Box$

We give, without proof, the following theorem, that is due to \cite{ren61}:
\begin{theorem}\label{RENY}
The logarithm is the only mathematical operation that satisfies:
\begin{itemize} 
\item Additivity: $I(m \times n) = I(m) + I(n)$, 
\item  Monotonicity: $I(m) \leq I(m+1)$ and 
\item Normalisation: $I(a)=1$.
\end{itemize}
\end{theorem}

By theorem \ref{RENY} the logarithm exactly represents the extensive qualities of a general notion of information in natural numbers. If functions are defined on natural numbers we can use the log function to measure the information: 
\[I(x)=\log x, I(f(x))=\log f(x)  \]

Combining theorem \ref{RENY} and lemma  \ref{EFF} we expect any computable bijection $f: \mathbb{N}^2 \rightarrow \mathbb{N}$ to be information efficient in the limit when measured in terms of the log operation. We cannot compute a double limit for the Cantor function $\pi^2$ directly, but we can compute the limit for  \emph{almost all} points by computing the infinite sets of limits on all lines $y=hx$, with $h>0$: 

\begin{lemma} \label{INEFF}
The cantor function $\pi^2(x,y) := 1/2(x + y)(x + y + 1)+y$ has in the limit constant information efficiency on all the lines $y = hx$ with $h>0$. 
\end{lemma}

Proof: We have to prove that $\lim_{x \rightarrow \infty} \delta(\pi^2(x,hx)) = c$. We compute the information efficiency in the limit on the line $y=hx$: 
  \begin{equation}
\lim_{x \rightarrow \infty} \delta(\pi^2(x,hx)) = 
\end{equation} 

\[
\lim_{x \rightarrow \infty} 
\log (\frac{1}{2} (x + h x + 1) ( x +  h x) +  h x)
- \log x - \log h x =
\]

\[
\log \lim_{x \rightarrow \infty} \frac
{\frac{1}{2} (x + h x + 1) ( x +  h x) +  h x}{hx^2} = \log  (\frac{1}{2h} + 1 +  \frac{h}{2}) = c
\]

$\Box$

We get insight as to why the Fueter - P\'{o}lya  conjecture is true, when we rewrite the third line of the proof above as: 

\[
\log \lim_{x \rightarrow \infty} \frac
{\pi^2(x,hx)}{hx^2} = c
\]

The term $hx^2$ in the fraction is generated by the fact that the input is two-dimensional. As a consequence no  function with order $>2$ can be information efficient: 

\begin{theorem} \label{NOHIGH}
There are no other polynomial functions  with degree $> 2$ that define a bijection between $\mathbb{N}$ and  $\mathbb{N}^2$.  
\end{theorem}
Proof: Suppose such a function $\pi^n$ of degree $n$ exists. Since it is a bijection it needs to be information conserving in the limit for the numbers on all  lines $y=hx$, with $h>0$. Consequently the conditions of lemma \ref{INEFF} hold:  $\lim_{x \rightarrow \infty} \delta(\pi^n(x,hx)) = c $, for al $h > 0$.This implies that  $\pi^n$  will stay asymptotically close to $\pi^2$ on all lines $y=hx$. Specifically, following the proof of lemma  \ref{INEFF}, for each line $y = hx$: 

\[
\log \lim_{x \rightarrow \infty} \frac
{\pi^n(x,hx)}{hx^2} = c
\]

The fact that $\pi^n$ has degree $n>2$ will not come to expression on any line $y=hx$, with $h>0$.  Consequently $\pi^n$ is not a function with degree higher than 2.  
$\Box$

\section{A version based on Kolmogorov complexity}
It is possible to rephrase the proof purely in terms of Kolmogorov complexity:

\begin{lemma}\label{EFFKOL}
Suppose A is an infinite set and $f :\mathbb{N} \rightarrow A$  and $f^{-1 }$ are computable bijections, then:
 
  \[\forall (x \in A) K(f^{-1}(x)) \leq K(x) + O(1) \]
  \[ \forall (y \in \mathbb{N}) K(f(y)) \leq K(y) + O(1)\] 
\end{lemma}
Proof: If we have $x$ and $f^{-1}$ is a program of finite length, then $K(f^{-1}(x)) = K(x) + K(f^{-1}) + c$.  If we have $y$ and $f$ is a program of finite length, then $K(f(y)) = K(y) + K(f) + c$. 
$\Box$

 Note that in terms of lemma \ref{EFFKOL} the logarithm is an upper bound for the Kolmogorov complexity of a natural number and that there is an infinite amount of cases for which the actual Kolmogorov complexity of a number is considerable smaller. We need to establish that the conditions of lemma  \ref{INEFF} also hold for a sufficient amount of numbers with an adequate distribution of cells in $\mathbb{N}^2$.

\begin{definition}
An infinite set of numbers  is \emph{typical in the limit} if it contains an infinite amount of incompressible numbers, e.g. there is a constant $c$ for which it contains an infinite subset for which the condition $K(x) > \log_2 x - c$ holds. 
\end{definition}
\begin{lemma}\label{DENS}
Any infinite set of natural numbers that has density $>0$ is typical in the limit. 
\end{lemma}
Proof: this is an immediate consequence of a well-known  counting argument from Kolmogorov complexity (\cite{LiVi08}): the density of the set of numbers compressible by more than a constant is $0$  in the limit, so any set with density $>0$ contains an infinite amount of incompressible numbers. $\Box$

We need to substitute these results in the proof of theorem \ref{NOHIGH}: Observe that for each line $y=hx$, for any $\epsilon > 0$ the neighbourhood $\{(x,y)| \  hx \leq  y \leq (h+ \epsilon)x\}$ in $\mathbb{N}^2$ is dense and thus typical in the limit. The fact that $\pi^n$ has degree $n>2$ will in the limit not come to expression on any neighbourhood $\{(x,y)| \ hx \leq  y \leq (h+ \epsilon)x\}$, with $h>0$.  Consequently  $\pi^n$ is not a function with degree higher than 2.

\section{Acknowledgements}
This research was partly supported by the Info-Metrics Institute of the American University in Washington, the Commit project of the Dutch science foundation NWO, the Netherlands eScience center and a Templeton Foundation’s Science and Significance of Complexity Grant supporting The Atlas of Complexity Project. I thank Peter Van Emde Boas, the editor and the anonymous referees for their insightful comments on earlier versions. 



\section{Bibliography}

\bibliographystyle{elsarticle-harv} 

\bibliography{adriaans}





\end{document}